\documentclass[12pt,fleqn]{book} 
\usepackage{tcolorbox}
\usepackage[top=1.0in,bottom=1.0in,left=1.0in,right=1.0in,headsep=10pt,letterpaper]{geometry} 
\usepackage{xcolor} 
\definecolor{ocre}{RGB}{52,177,201} 

\usepackage{avant} 
\usepackage{mathptmx} 

\usepackage{microtype} 
\usepackage[utf8]{inputenc} 
\usepackage[T1]{fontenc} 

\usepackage{titlesec} 

\usepackage{graphicx} 
\graphicspath{{Pictures/}{figures/cosmo/}} 

\usepackage{lipsum} 

\usepackage{tikz} 

\usepackage[english]{babel} 

\usepackage{enumitem} 
\setlist{nolistsep} 

\usepackage{booktabs} 

\usepackage{eso-pic} 

\usepackage{titletoc} 

\contentsmargin{0cm} 
\titlecontents{chapter}[1.25cm] 
{\addvspace{15pt}\large\sffamily\bfseries} 
{\color{ocre!60}\contentslabel[\Large\thecontentslabel]{1.25cm}\color{ocre}} 
{}
{\color{ocre!60}\normalsize\sffamily\bfseries\;\titlerule*[.5pc]{.}\;\thecontentspage} 
\titlecontents{section}[1.25cm] 
{\addvspace{5pt}\sffamily\bfseries} 
{\contentslabel[\thecontentslabel]{1.25cm}} 
{}
{\sffamily\hfill\color{black}\thecontentspage} 
[]
\titlecontents{subsection}[1.25cm] 
{\addvspace{1pt}\sffamily\small} 
{\contentslabel[\thecontentslabel]{1.25cm}} 
{}
{\sffamily\;\titlerule*[.5pc]{.}\;\thecontentspage} 
[]

\titlecontents{lsection}[0em] 
{\footnotesize\sffamily} 
{}
{}
{}

\titlecontents{lsubsection}[.5em] 
{\normalfont\footnotesize\sffamily} 
{}
{}
{}

\usepackage{fancyhdr} 

\fancypagestyle{plain}{\fancyhead{}} 

\makeatletter
\renewcommand{\cleardoublepage}{
\clearpage\ifodd\c@page\else
\hbox{}
\vspace*{\fill}
\thispagestyle{empty}
\newpage
\fi}

\usepackage{amsmath,amsfonts,amssymb,amsthm} 

\newtheoremstyle{ocrenumbox}
{0pt}
{0pt}
{\normalfont}
{}
{\small\bf\sffamily\color{ocre}}
{\;}
{0.25em}
{\small\sffamily\color{ocre}\thmname{#1}\nobreakspace\thmnumber{\@ifnotempty{#1}{}\@upn{#2}}
\thmnote{\nobreakspace\the\thm@notefont\sffamily\bfseries\color{black}---\nobreakspace#3.}} 

\newtheoremstyle{blacknumex}
{5pt}
{5pt}
{\normalfont}
{} 
{\small\bf\sffamily}
{\;}
{0.25em}
{\small\sffamily{\tiny\ensuremath{\blacksquare}}\nobreakspace\thmname{#1}\nobreakspace\thmnumber{\@ifnotempty{#1}{}\@upn{#2}}
\thmnote{\nobreakspace\the\thm@notefont\sffamily\bfseries---\nobreakspace#3.}}

\newtheoremstyle{blacknumbox} 
{0pt}
{0pt}
{\normalfont}
{}
{\small\bf\sffamily}
{\;}
{0.25em}
{\small\sffamily\thmname{#1}\nobreakspace\thmnumber{\@ifnotempty{#1}{}\@upn{#2}}
\thmnote{\nobreakspace\the\thm@notefont\sffamily\bfseries---\nobreakspace#3.}}

\newtheoremstyle{ocrenum}
{5pt}
{5pt}
{\normalfont}
{}
{\small\bf\sffamily\color{ocre}}
{\;}
{0.25em}
{\small\sffamily\color{ocre}\thmname{#1}\nobreakspace\thmnumber{\@ifnotempty{#1}{}\@upn{#2}}
\thmnote{\nobreakspace\the\thm@notefont\sffamily\bfseries\color{black}---\nobreakspace#3.}} 
\makeatother

\newcounter{dummy}
\numberwithin{dummy}{section}
\theoremstyle{ocrenumbox}
\newtheorem{theoremeT}[dummy]{Theorem}

\newtheorem{exerciseT}{Exercise}[chapter]
\theoremstyle{blacknumex}
\newtheorem{exampleT}{Example}[chapter]
\theoremstyle{blacknumbox}

\newtheorem{definitionT}{Definition}[section]
\newtheorem{corollaryT}[dummy]{Corollary}
\theoremstyle{ocrenum}

\RequirePackage[framemethod=default]{mdframed} 

\newmdenv[skipabove=7pt,
skipbelow=7pt,
backgroundcolor=black!5,
linecolor=ocre,
innerleftmargin=5pt,
innerrightmargin=5pt,
innertopmargin=5pt,
leftmargin=0cm,
rightmargin=0cm,
innerbottommargin=5pt]{tBox}

\newmdenv[skipabove=7pt,
skipbelow=7pt,
rightline=false,
leftline=true,
topline=false,
bottomline=false,
backgroundcolor=ocre!10,
linecolor=ocre,
innerleftmargin=5pt,
innerrightmargin=5pt,
innertopmargin=5pt,
innerbottommargin=5pt,
leftmargin=0cm,
rightmargin=0cm,
linewidth=4pt]{eBox}	

\newmdenv[skipabove=7pt,
skipbelow=7pt,
rightline=false,
leftline=true,
topline=false,
bottomline=false,
linecolor=ocre,
innerleftmargin=5pt,
innerrightmargin=5pt,
innertopmargin=0pt,
leftmargin=0cm,
rightmargin=0cm,
linewidth=4pt,
innerbottommargin=0pt]{dBox}	

\newmdenv[skipabove=7pt,
skipbelow=7pt,
rightline=false,
leftline=true,
topline=false,
bottomline=false,
linecolor=gray,
backgroundcolor=black!5,
innerleftmargin=5pt,
innerrightmargin=5pt,
innertopmargin=5pt,
leftmargin=0cm,
rightmargin=0cm,
linewidth=4pt,
innerbottommargin=5pt]{cBox}

\makeatletter
\renewcommand{\@seccntformat}[1]{\llap{\textcolor{ocre}{\csname the#1\endcsname}\hspace{1em}}}
\renewcommand{\section}{\@startsection{section}{1}{\z@}
{-2ex \@plus -1ex \@minus -.2ex}
{1ex \@plus.1ex }
{\normalfont\large\sffamily\bfseries}}
\renewcommand{\subsection}{\@startsection {subsection}{2}{\z@}
{-2ex \@plus -0.1ex \@minus -.2ex}
{0.5ex \@plus.2ex }
{\normalfont\sffamily\bfseries}}
\renewcommand{\subsubsection}{\@startsection {subsubsection}{3}{\z@}
{-2ex \@plus -0.1ex \@minus -.2ex}
{.2ex \@plus.2ex }
{\normalfont\small\sffamily\bfseries}}
\renewcommand\paragraph{\@startsection{paragraph}{4}{\z@}
{-2ex \@plus-.2ex \@minus .2ex}
{.1ex}
{\normalfont\small\sffamily\bfseries}}

\usepackage{hyperref}
\hypersetup{hidelinks,backref=true,pagebackref=true,hyperindex=true,colorlinks=false,breaklinks=true,urlcolor= ocre,bookmarks=true,bookmarksopen=false,pdftitle={Title},pdfauthor={Author}}

\newcommand{\thechapterimage}{}
\newcommand{\chapterimage}[1]{\renewcommand{\thechapterimage}{#1}}

\def\thechapter{\arabic{chapter}}
\def\@makechapterhead#1{
\thispagestyle{empty}
{\centering \normalfont\sffamily
\ifnum \c@secnumdepth >\m@ne
\if@mainmatter
\startcontents
\begin{tikzpicture}[remember picture,overlay]
\node at (current page.north west)
{\begin{tikzpicture}[remember picture,overlay]
\node[anchor=north west,inner sep=0pt] at (0,0) {\includegraphics[width=\paperwidth]{\thechapterimage}};
\draw[anchor=west] (5cm,-9cm) node [rounded corners=20pt,fill=ocre!10!white,text opacity=1,draw=ocre,draw opacity=1,line width=1.5pt,fill opacity=.6,inner sep=12pt]{\huge\sffamily\bfseries\textcolor{black}{\thechapter. #1\strut\makebox[22cm]{}}};
\end{tikzpicture}};
\end{tikzpicture}}
\par\vspace*{230\p@}
\fi
\fi}

\def\@makeschapterhead#1{
\thispagestyle{empty}
{\centering \normalfont\sffamily
\ifnum \c@secnumdepth >\m@ne
\if@mainmatter
\begin{tikzpicture}[remember picture,overlay]
\node at (current page.north west)
{\begin{tikzpicture}[remember picture,overlay]
\node[anchor=north west,inner sep=0pt] at (0,0) {\includegraphics[width=\paperwidth]{\thechapterimage}};
\draw[anchor=west] (5cm,-6cm) node [rounded corners=20pt,fill=ocre!10!white,fill opacity=.6,inner sep=12pt,text opacity=1,draw=ocre,draw opacity=1,line width=1.5pt]{\LARGE\sffamily\bfseries\textcolor{black}{#1\strut\makebox[22cm]{}}};
\end{tikzpicture}};
\end{tikzpicture}}
\par\vspace*{130\p@}
\fi
\fi
}
\makeatother

\usepackage{graphicx}
\graphicspath{{./figures/}}
\usepackage{appendix}
\usepackage{latexsym,amsmath,amsfonts,amssymb,booktabs}
\usepackage[font=small]{caption}
\usepackage{slashed,upgreek,amscd,cancel,tensor,color}
\usepackage{adjustbox}
\usepackage[numbers,sort&compress,square]{natbib}
\usepackage{epsfig,latexsym}
\usepackage{url}
\numberwithin{equation}{section}
\usepackage{doi}
\usepackage{subcaption}
\usepackage{mathtools}
\usepackage{upgreek}
\usepackage{stfloats}
\usepackage{afterpage}
\usepackage{multirow}

\begin{document}


\chapterimage{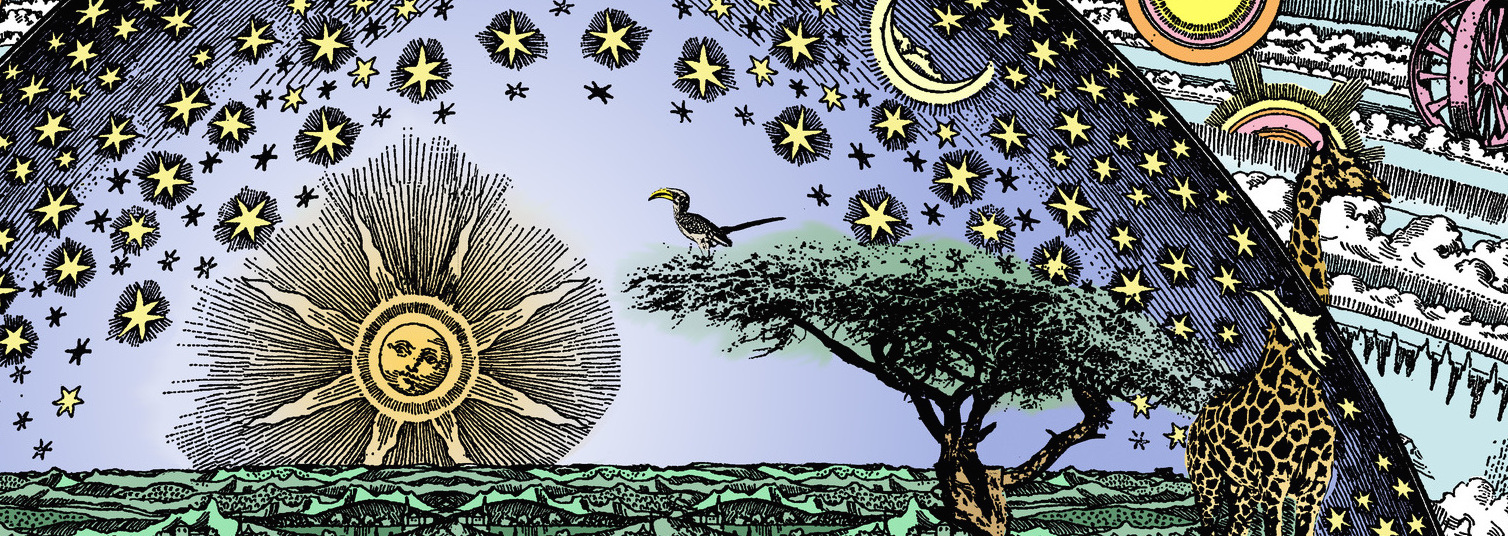} 
\chapter*{\Large Cosmology and the Early Universe}
\begin{center}
\Large
\textbf{Astro2020 Science White Paper} \linebreak


COSMOLOGY AND THE EARLY UNIVERSE 
\end{center} 

\normalsize


\noindent \textbf{Thematic Areas:} 
\begin{itemize}
\item Cosmology and Fundamental Physics 
\item Multi-Messenger Astronomy and Astrophysics
\end{itemize}


\vspace{0.75cm}

\noindent \textbf{Principal Author}: 
\newline
\vspace{-0.5cm}

 Name:	B.S. Sathyaprakash
 \newline
\vspace{-0.5cm}

Institution: The Pennsylvania State University
 \newline
\vspace{-0.5cm}

Email: bss25@psu.edu
 \newline
\vspace{-0.5cm}

Phone: +1-814-865-3062 
 \newline

 \noindent 
\textbf{Lead Co-authors:} 
Enis Belgacem (University of Geneva),
Daniele Bertacca (Universit\'a degli Studi di Padova),
Chiara Caprini (Universit\'e Paris-Diderot, France),
Giulia Cusin (University of Oxford, United Kingdom),
Yves Dirian (University of Geneva and University of Zuerich),
Xilong Fan (Wuhan University and Hubei University of Education, China),
Daniel Figueroa (\'Ecole Polytechnique F\'ed\'erale de Lausanne, Switzerland),
Stefano Foffa (University of Geneva),
Evan Hall (MIT),
Jan Harms (Gran Sasso Science Institute, Italy), 
Michele Maggiore (University of Geneva, Switzerland),
Vuk Mandic (University of Minnesota), 
Andrew Matas (Max Planck Institute for Gravitational Physics, Potsdam, Germany),
Tania Regimbau (Universit\'e Grenoble Alpes, France),
Mairi Sakellariadou (Kings College London, United Kingdom), 
Nicola Tamanini (Max Planck Institute for Gravitational Physics, Potsdam), 
Eric Thrane (Monash University, Australia)
\newline

\noindent 
Click here for \href{https://docs.google.com/spreadsheets/d/1uNKEW77Fm-_nc21_3jSOX-P4ecRxsxA5UgDOD4bkG9Y/edit#gid=1617475973}{\bf other co-authors and supporters}




\clearpage
\subsection*{Cosmology and the Early Universe}

Gravity assembles structures in the Universe from the smallest scale of planets to galaxy clusters and the largest scale of the Universe itself. Yet, until recently gravity played only a passive role in observing the Universe. Almost everything we know about the universe, from its hot primeval phase to recent accelerated expansion, the most powerful explosions such as hypernovae and brightest objects like quasars, comes from electromagnetic waves. Important exceptions include cosmic rays and neutrinos that have allowed to probe complementary phenomena.

September 14, 2015 forever changed the role of gravity in our exploration of the Universe, when the Laser Interferometer Gravitational-wave Observatory (LIGO) observed gravitational waves (GWs) from a pair of coalescing black holes \cite{Abbott:2016blz}. Since then LIGO and the Virgo detector have observed nine more binary black hole coalescence events \cite{LIGOScientific:2018mvr}. Moreover, the joint discovery on August 17, 2017 of a binary neutron star merger by LIGO, Virgo \cite{TheLIGOScientific:2017qsa} and the space-based Fermi and INTEGRAL gamma ray detectors \cite{Monitor:2017mdv} triggered an electromagnetic follow-up of the source and generated a treasure trove of data that has helped ushered in a new era in cosmology \cite{Abbott:2017xzu}.

\begin{tcolorbox}[standard jigsaw,colframe=ocre,colback=ocre!10!white,opacityback=0.6,coltext=black]
Future gravitational-wave observations will enable exploration of {\em particle physics, early universe,} and {\em cosmology,} that are among the core topics of the Thematic Area 7 of Astro-2020:
\begin{itemize}[leftmargin=*]
\item {\bf Early Universe.} What evolutionary phases took place in the early Universe, their energy scales? How did they transition from one into another and the Universe observed today?
\item {\bf Dark Matter.} What is the nature of dark matter and what was its role in structure formation? Do primordial black holes account for some fraction of dark matter?
\item {\bf Standard Siren Cosmology.} Is dark energy just a cosmological constant, or does the dark energy equation-of-state vary with redshift?
\item {\bf Modified Theories of Gravity.} Do gravitational waves propagate from their sources in the same way as electromagnetic waves do? How do extra dimensions and other modified theories of gravity affect the propagation of gravitational waves from their sources?
\end{itemize}
\end{tcolorbox}

Merging binaries of neutron stars and black holes are self-calibrating standard sirens \cite{Schutz:1986gp}. General relativity completely determines the time evolution of the amplitude (which gives their apparent luminosity) and frequency (which is a measure of their absolute luminosity) of emitted gravitational waves. Thus, distances to host galaxies can be inferred without the aid of the cosmic distance ladder, thereby providing a new tool for cosmology. Moreover, the inferred merger rates of binary black holes and neutron stars imply that future detectors will observe a stochastic background formed from the astrophysical population of binaries at cosmological distances \cite{TheLIGOScientific:2016htt,gw170817_stoch}; such observations will reveal the history of star formation from a time when the universe was still assembling its first stars and galaxies. Buried under this background could be stochastic signals of primordial origin that could provide insights into the physics and energy scales of the earliest evolutionary phases in the history of the Universe---scales that are not accessible to current or planned particle physics experiments (see, e.g., \cite{Grishchuk:1974ny,Starobinsky:1979ty,Rubakov:1982df,Fabbri:1983us,Sorbo:2011rz,Barnaby:2010vf,Giovannini:1999bh,Giovannini:1998bp,
Boyle:2005se,Boyle:2007zx,Liu:2015psa,Figueroa:2016dsc,Caprini:2018mtu,Kosowsky:1991ua,Kosowsky:1992vn,Kamionkowski:1993fg,Huber:2008hg,Weir:2016tov,
Hindmarsh:2013xza,Hindmarsh:2015qta,Hindmarsh:2017gnf,Caprini:2009yp,Kahniashvili:2008pe,Kahniashvili:2008er}).

\begin{figure*}
    \centering
    \includegraphics[width=0.45\textwidth]{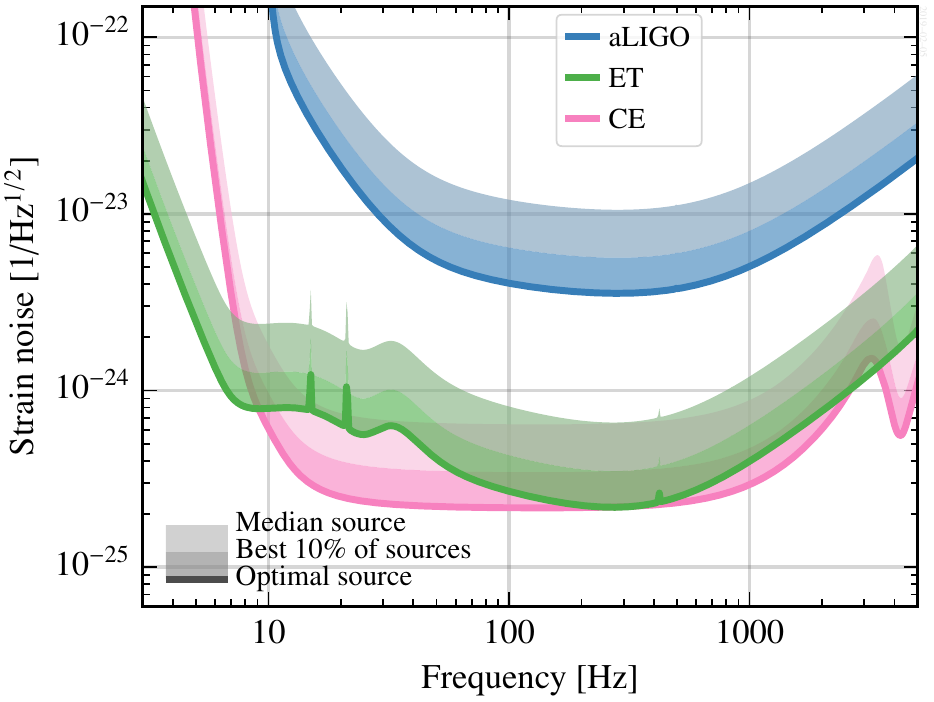}
    \includegraphics[width=0.45\textwidth]{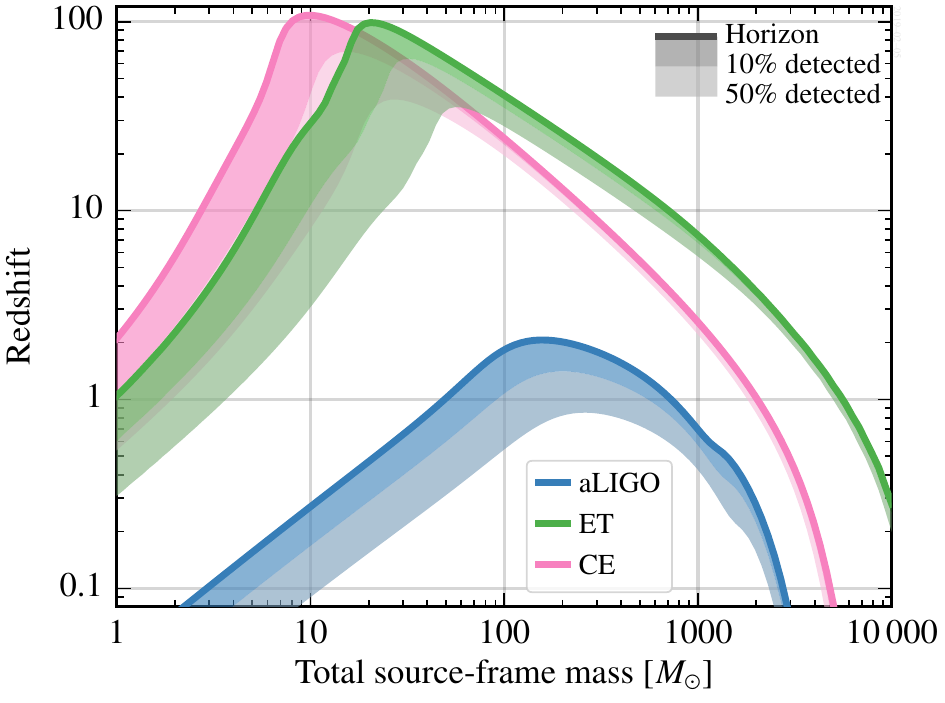}
    \caption{\small Strain noise (left) for monochromatic sources distributed isotropically in sky location and inclination and
    astrophysical reach (right) for equal-mass, nonspinning binaries similarly distributed.}
    \label{fig:noises_percentiles}
    \vskip-0.5cm
\end{figure*}

Advanced LIGO and Virgo will only be first steps in this new endeavor that is guaranteed to change our perception of the Universe in the coming decades. Indeed, the next generation of GW observatories, such as the Einstein Telescope and Cosmic Explorer (referred to as 3G), will witness merging black holes and neutron stars when the Universe was still in its infancy assembling its first stars and black holes
as shown in Figure \ref{fig:noises_percentiles}. At such sensitivity levels we can expect to witness extremely bright events that could reveal subtle signatures of new physics. 3G observatories promise to deliver data that could transform the landscape of physics, addressing some of the most pressing problems in fundamental physics and cosmology \cite{Sathyaprakash:2009xs}.

\subsection*{Early Universe}
\begin{figure}[t]
\centering
\includegraphics[width=0.80\textwidth]{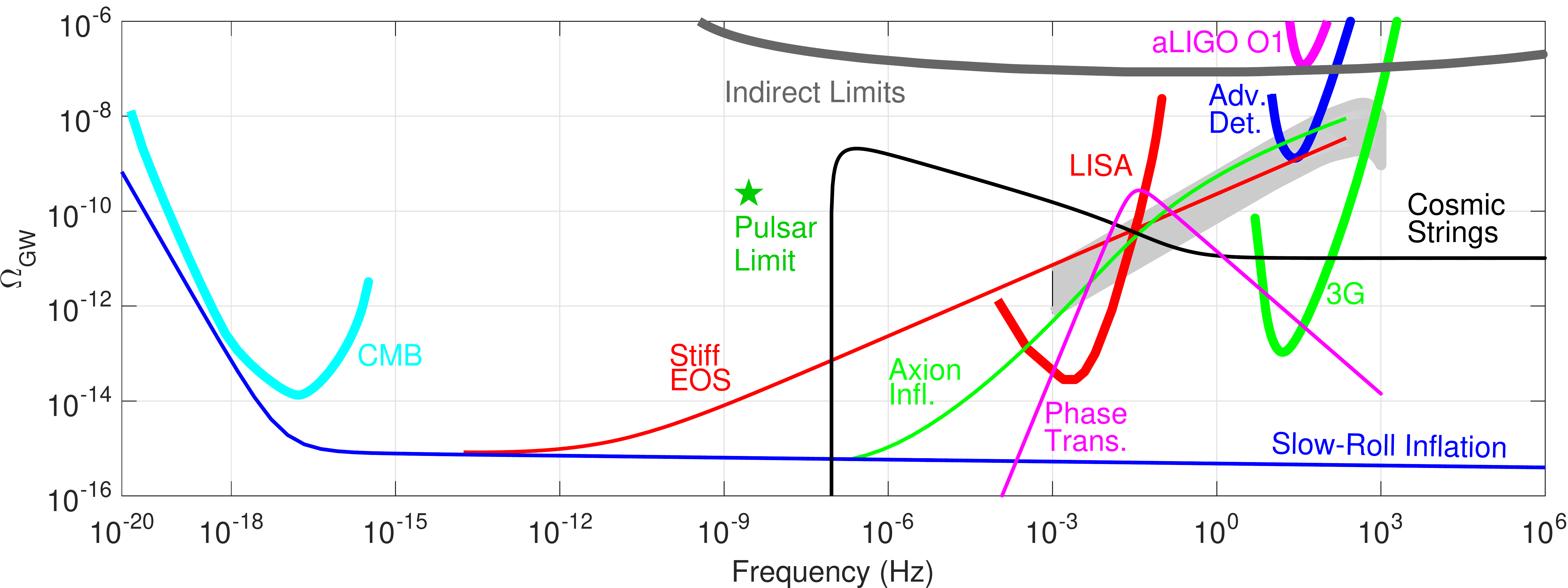}
\caption{\small Stochastic gravitational-wave background for
several proposed model spectra in comparison with past
measurements (Advanced LIGO upper limit
\cite{TheLIGOScientific:2016dpb}, constraints based on the
big bang nucleosynthesis and CMB observations
\cite{Lasky:2015lej}, low-$l$ CMB observations
\cite{Lasky:2015lej}, and pulsar timing
\cite{Lasky:2015lej}),  and future expected sensitivities
(the final sensitivity of Advanced LIGO
\cite{PhysRevLett.116.131102}, Cosmic Explorer
\cite{Abbott:2017re}, and LISA, all assuming 1 year of
exposure \cite{LisaProposal:2017,Axen:2018zvb}. The gray
band denotes the expected amplitude of the background due to
the cosmic population of compact binary mergers, based on
the observed BBH and BNS systems
\cite{2018PhRvL.120i1101A}.}
\label{fig:landscape}
    \vskip-0.5cm
\end{figure}

Gravitational wave observations may inform us about the history and structure of the Universe in at least two ways: by studies of individual sources at cosmological distances that give information about its geometry and kinematics and by direct observation of a stochastic GW background. In turn, a stochastic background could either be astrophysical in origin, generated by any of a myriad of astrophysical systems or it could come from the Big Bang itself, generated by quantum processes associated with inflation or with spontaneous symmetry breaking in the early Universe. Figure \ref{fig:landscape} shows examples of energy density spectra for some of the cosmological background models in comparison with the best current upper limits and future expected detector sensitivities. Cosmological background is probably the most fundamentally important observation that GW detectors can make. However, the astrophysical background may mask the primordial background over much of the accessible spectrum, while still carrying important information about the evolution of structure in the Universe. Techniques are being developed to identify and estimate these various contributions to the stochastic GW background.
%
%

\noindent {\bf Inflation:} 
Cosmological stochastic background is expected to be produced during inflation via the mechanism of amplification of vacuum fluctuations. In the standard slow-roll inflationary model this background is likely below the sensitivity of 3G detectors \cite{Grishchuk:1974ny,Starobinsky:1979ty,Rubakov:1982df,Fabbri:1983us}. However, additional processes during or immediately after inflation could lead to significant amplification of the stochastic background in the frequency band of 3G detectors.  For example, efficient gauge field excitation due to axionic coupling to the inflaton could extend inflation and boost gravitational-wave production with a significant blue tilt~\cite{Sorbo:2011rz,Barnaby:2010vf,Cook:2011hg,Cook:2013xea,Bartolo:2016ami}. Similar large blue tilts in the GW energy spectrum are also possible if inflation is followed by another (presently unknown) phase with a stiff equation of state ($w > 1/3$), again resulting in possible detectability with 3G detectors~\cite{Giovannini:1999bh,Giovannini:1998bp,Boyle:2005se,Boyle:2007zx,Liu:2015psa,Figueroa:2016dsc}.

\noindent {\bf First order phase transitions:} 
Following the end of inflation, the Universe has undergone the QCD, the electroweak, and potentially other phase transitions.
Currently, an experimentally verified physical model is lacking for energy scales higher than the electroweak one, where several proposed extensions of the Standard Model of Particles predict the occurrence of phase transitions. Any experimental confirmation that such phase transitions took place in the early Universe would therefore constitute an invaluable piece of information for our understanding of particle theory underlying the Universe at very high energies. To be an efficient direct source of GWs, a phase transition must be of first order. First order phase transitions proceed through the nucleation of bubbles of the true vacuum, energetically more favourable, in the space-filling false vacuum.

The dynamics of the bubble expansion and collision is phenomenologically very rich, and the sources of GWs are the tensor anisotropic stresses generated by these multiple phenomena: the bubble wall’s expansion~\cite{Kosowsky:1991ua,Kosowsky:1992vn,Kamionkowski:1993fg,Huber:2008hg,Weir:2016tov}, the sound waves in the plasma~\cite{Hindmarsh:2013xza,Hindmarsh:2015qta,Hindmarsh:2017gnf}, and the subsequent magnetohydrodynamic turbulence~\cite{Caprini:2009yp,Kahniashvili:2008pe,Kahniashvili:2008er}. 

\noindent {\bf Cosmic Strings:} If the broken vacuum manifold is topologically nontrivial, topological defects such as cosmic strings may arise in the aftermath of a phase transition. 
Cosmic strings are expected to form in the context of Grand Unified Theories applied in the early Universe \cite{Jeannerot:2003qv}. For Nambu-Goto strings the string tension is the only free parameter; it defines the energy scale of the phase transition accompanied by the spontaneous symmetry breaking scale
leading to the cosmic strings formation. It is also possible to form a network of fundamental cosmic (super)strings, in which case the network is also affected by the probability that two strings would reconnect when they intersect.

Cosmic strings predominantly decay by the formation of loops and the subsequent GW emission by cosmic string cusps and kinks \cite{Vachaspati:1984gt,Sakellariadou:1990ne}. Searches for both individual bursts of GWs due to cosmic string cusps and for the stochastic background due to cusps and kinks have been conducted using data from Advanced LIGO and Advanced Virgo, placing a strong constraint on the string tension for the three known Nambu-Goto string models
\cite{Blanco-Pillado:2017oxo,Ringeval:2017eww,Abbott:2017mem,Jenkins:2018lvb}, $G\mu\lesssim10^{-11}$. The 3G detectors will be able to substantially improve on these bounds, by as much as 8 orders of magnitude relative to current detectors, depending on the model~(see Fig.~\ref{fig:landscape}).

\subsection*{Dark Matter} 
While there is overwhelming evidence supporting existence of dark matter in the Universe, its composition is currently unknown. Numerous dark matter models have been proposed, some of which predict gravitational-wave signatures accessible to 3G detectors. 

\noindent {\bf Primordial Black Holes:} Black holes formed in high-density regions in the early universe could gravitationally interact to form stellar-mass compact binaries, which would then inspiral and merge, emitting GWs in the process. Such binaries could form via GW emission during the encounter of two black holes \cite{PhysRevLett.116.201301}, or via tidal forces due to neighboring black holes \cite{PhysRevLett.117.061101}. The primordial binary black hole population would evolve differently with redshift than the stellar population, which would also result in a different stochastic background energy density spectrum~\cite{PhysRevLett.117.201102}. These effects will be accessible to 3G detectors, enabling them to establish whether primordial black holes make a significant contribution to dark matter.

\noindent {\bf Dark Matter and Binary Dynamics:} Binary black holes evolving in a dark-matter rich environment will not only accrete the surrounding material, but also exert a gravitational drag on the dark matter medium, which affects the inspiral dynamics. This drag and accretion effect accumulated over a large number of orbits could be detected by a combination of observatories in space and 3G detectors. Dark matter could also accumulate in a finite-size core inside a star via interactions with Standard Model particles, which might imprint a signature on the merger phase of two of these objects or even lead to formation of light black holes inside the dense interior of neutron stars. These effects could be observed by 3G detectors.

\noindent {\bf Dark Photons:} A dark photon is proposed to be a light but massive gauge boson of a U(1) extension of the Standard Model. If sufficiently light, the local occupation number of the dark photon could be much larger than one, so it can then be treated as a coherently oscillating background field that imposes an oscillating force on objects that carry dark charge. The oscillation frequency is determined by the mass of the dark photon. Such effects could result in a stochastic background that could be measured by 3G detectors, potentially exploring large parts of the parameter space of such models~\cite{PhysRevLett.121.061102}.

\subsection*{Astrophysical Binary Foregrounds}

The cosmological population compact binary mergers will give
rise to a stochastic foreground of GWs
\cite{Regimbau:2016ike, Mandic:2016lcn,
Dvorkin:2016okx, Nakazato:2016nkj, Dvorkin:2016wac,
Evangelista:2014oba}, potentially masking the cosmological
backgrounds discussed above. Figures \ref{fig:landscape}
shows that the amplitude of this foreground is several
orders of magnitude stronger than most cosmological
backgrounds. However, a large fraction of the binary merger
signals will be individually detected by the 3G detector
network, allowing for the possibility to subtract this
foreground. 

In principle, it may be possible to subtract binary black
hole events observed by 3G detectors to probe a cosmic
background of $\Omega_{\rm GW}\sim 10^{-13}$
\citep{Regimbau:2016ike}. However, small errors in subtraction
could lead to a substantial residual foreground. Moreover,
binary neutron star signals will be more difficult to remove, 
given their longer duration, and the fact that on
average many signals will overlap in time. 

Astrophysical sources other than compact binary mergers
could also contribute to the astrophysical background
including isolated neutron stars \cite{Surace:2015ppq,
Talukder:2014eba, Lasky:2013jfa}, stellar core collapse
\cite{Crocker:2017agi, Crocker:2015taa} and population III
binaries~\cite{Kowalska:2012ba}.  The study of the
properties of the astrophysical background will allow us to
put new types of constraints on astrophysical models and can
impact astrophysics as much as cosmic microwave background
did for cosmology during the past decades. In particular, it
will be possible to extract precise data on galactic and
stellar physics and their evolution and on astrophysical
processes of GW emission.

\subsection*{Dark energy and cosmological parameters}
Dark energy has been studied by investigating its effect on the expansion of the universe, specifically by probing the relationship between the luminosity distance and redshift of ``standard candles'', such as Type Ia supernovae. Gravitational waves offer another approach to this problem, using compact binary mergers as ``standard sirens''. The GW signal from coalescing binaries allows a direct measurement of their luminosity distance $d_L$~\cite{Schutz:1986gp}.
Dark energy can then be probed either by using redshift measurements from electromagnetic counterparts, or by using statistical methods~\cite{Dalal:2006qt,MacLeod:2007jd,Nissanke:2009kt,Cutler:2009qv,Sathyaprakash:2009xt,Zhao:2010sz,DelPozzo:2011yh,
Nishizawa:2011eq,Taylor:2012db,Camera:2013xfa,Tamanini:2016zlh,Caprini:2016qxs,Cai:2016sby,Belgacem:2017ihm,Belgacem:2018lbp}.
3G detectors will measure standard sirens up to large redshifts, $z\, \sim 10$, significantly farther than what is possible with the standard candles, while never having to rely on the cosmological distance ladder. Studies have shown that measuring cosmological parameters with $1-2\%$ uncertainties will be possible, for example by using 1000 observed binary neutron stars with GRB counterparts \cite{Belgacem:2017ihm,Belgacem:2018lbp}. This level of uncertainty is comparable to what is achievable with electromagnetic measurements, but is susceptible to a completely different set of systematic errors from those due, e.g., to supernovae, offering an independent measure.

\begin{figure}[t]
\centering
\includegraphics[width=0.45\textwidth]{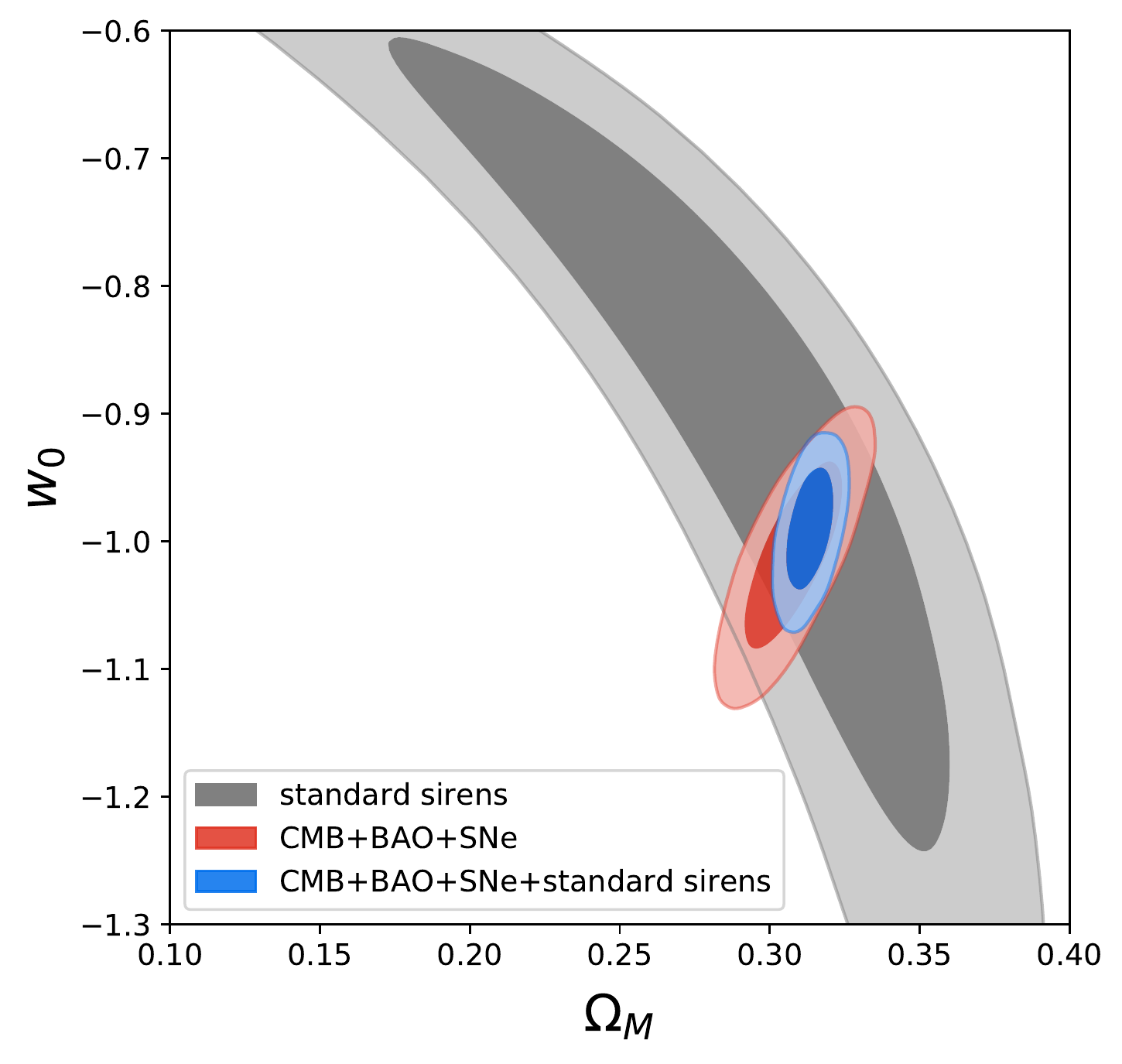}\hspace{10mm}
\includegraphics[width=0.45\textwidth]{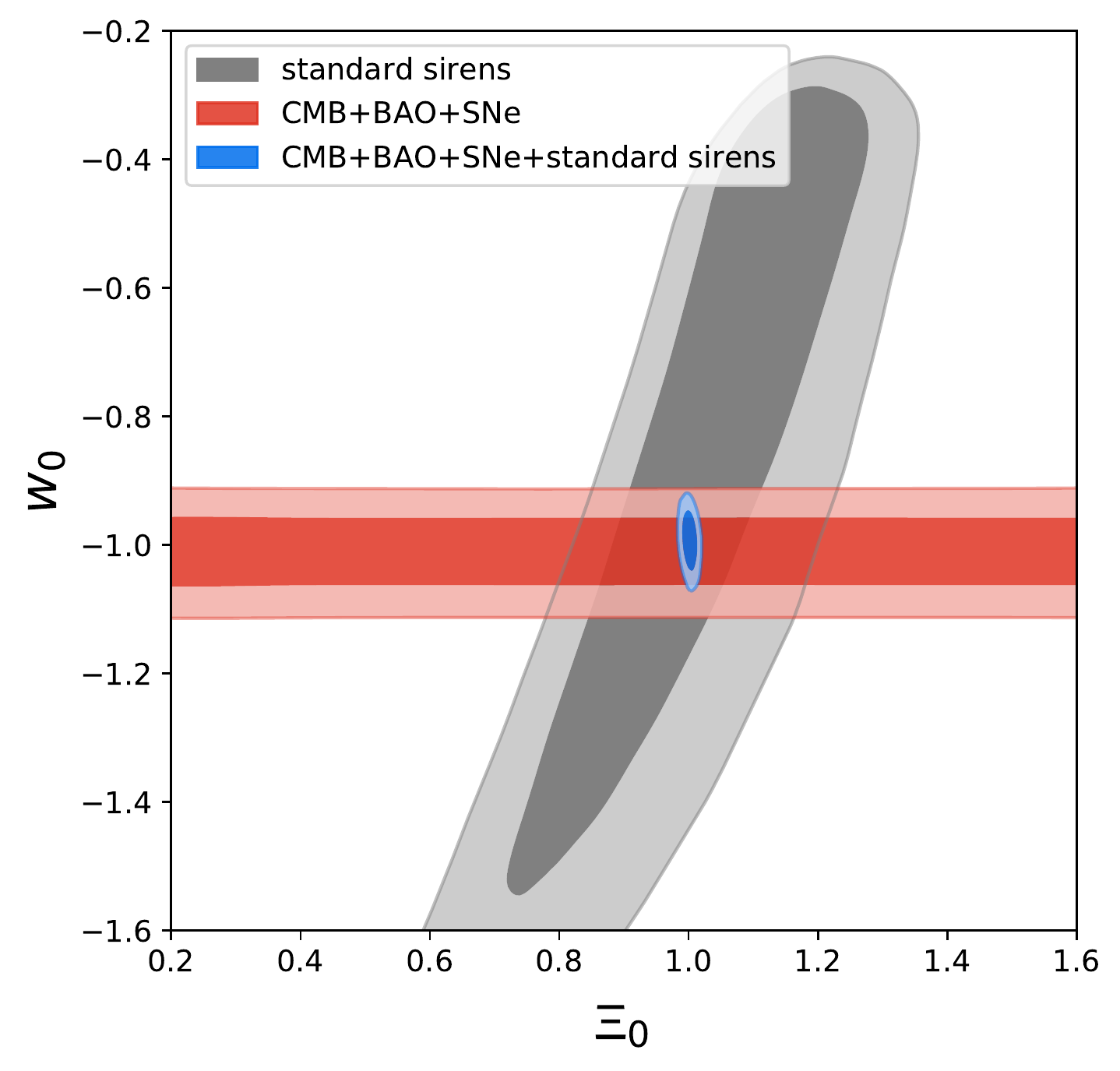}
\caption{\small Left: the  $1\sigma$ and $2\sigma$
contours  of the two-dimensional likelihood in the $(\Omega_\mathrm{M},w_0)$ plane with the combined  contribution from {\em Planck}
CMB data + BAO + SNe (red), the contribution from  $10^3$ standard sirens at ET (gray), and the total combined result (blue). Right:
the  two-dimensional likelihood in the $(\Xi_0,w_0)$ plane.  The mock catalog of standard sirens is generated  assuming  $\Lambda$CDM as the fiducial model ($w_0=-1$ and $\Xi_0=1$) and the plot shows the accuracy in the reconstruction of of $w_0$ and $\Xi_0$ (plot from \cite{Belgacem:2018lbp}).}
\label{fig:w0wa}
    \vskip-0.5cm
\end{figure}

Standard sirens are sensitive to another powerful signature of the dark energy sector that is not accessible to electromagnetic observations. A generic modified gravity theory induces modifications with respect to the standard model of cosmology both in the cosmological background evolution and in the cosmological perturbations.
Indeed, theories with extra dimensions~\cite{Deffayet:2007kf}, some scalar-tensor theories of the Horndeski class (including Brans-Dicke)~\cite{Saltas:2014dha,Lombriser:2015sxa,Arai:2017hxj,Amendola:2017ovw,Linder:2018jil}, as well as a nonlocal modification of gravity~\cite{Maggiore:2013mea,Maggiore:2014sia,Belgacem:2017cqo,Belgacem:2017ihm,Belgacem:2018lbp}, are characterized by GWs propagating at the speed of light but with their amplitude decreasing differently with the scale factor than in GR. Consequently, the standard sirens would measure a ``GW luminosity distance'' as opposed to the standard ``electromagnetic'' luminosity distance. The 3G detectors will be sufficiently sensitive to search for this deviation, and thereby probe multiple classes of modified theories of gravity in the context of their Dark Energy content~\cite{Belgacem:2017ihm,Belgacem:2018lbp}. Modified gravity theories also induce modified cosmological perturbations in the scalar sector, which then leave measurable signatures via the Integrated-Sachs Wolfe (ISW) effect~\cite{Laguna:2009re},  the effect of the ISW on a primordial stochastic background~\cite{Contaldi:2016koz, Cusin:2018rsq, Cusin:2017mjm, Cusin:2017fwz}, the effect of peculiar velocities \cite{Kocsis:2005vv,Bonvin:2016qxr}, and of lensing~\cite{Holz:2005df, Hirata:2010ba, Dai:2016igl, Dai:2017huk}. 3G detectors may also be sensitive to these effects for some of the modified gravity theories~\cite{Bertacca:2017vod}.

\clearpage

\chapterimage{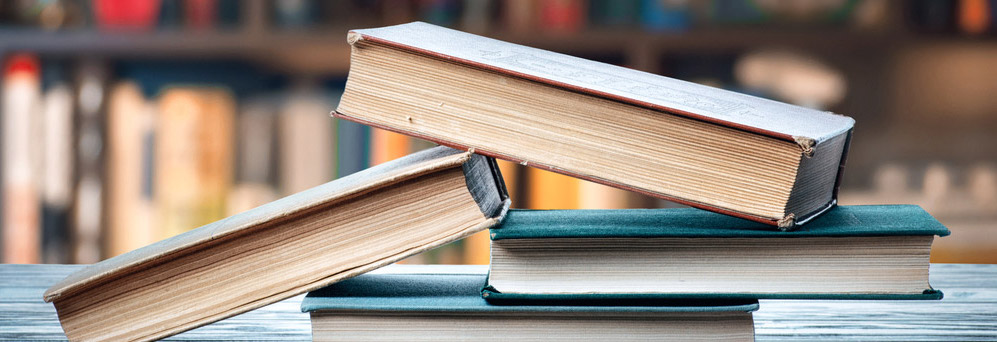} 
\bibliographystyle{utphys}
\bibliography{wp-all,3g,wp}
\end{document}